# REALISTIC INTERPRETATION OF ENTANGLED STATE: A DEFENSE AND APPLICATION TO HARDY'S EXPERIMENT

Boon Leong Lan

*School of Engineering, Monash University*
*46150 Bandar Sunway, Selangor, Malaysia*
email: lan.boon.leong@eng.monash.edu.my

**Abstract**

Two criticisms which have prevented the realistic interpretation of entangled state from being widely accepted are addressed and shown to be unfounded. A local realistic theory, which reproduces all the quantum probabilistic predictions, is constructed for Hardy's experiment based on the realistic interpretation of the entangled two-particle Hardy state.

## 1. Introduction

Consider a system of two spin-½ particles which is mathematically described by an entangled state such as this superposition of states:

$$\frac{1}{\sqrt{2}}|+z\rangle_1|+z\rangle_2 + \frac{1}{\sqrt{2}}|-z\rangle_1|-z\rangle_2. \qquad (1)$$

In the orthodox interpretation [1-3] of entangled state, the system of two particles does not have any definite state before measurement. Measurement on one particle compels that particle to acquire a definite state and instantaneously triggers the other particle, which is spatially separated from the first, to also acquire a definite state. For the entangled state in Eq. (1), if the spin z measurement on particle 1 induces it to acquire state $|+z\rangle_1$, particle 2 is triggered to acquire state $|+z\rangle_2$; if particle 1 is induced to acquire state $|-z\rangle_1$ instead, particle 2 is triggered to acquire state $|-z\rangle_2$. In contrast, in the realistic interpretation [3-5] of entangled state, the system of two particles has definite states before measurement, one state for each possible joint measurement. A chosen joint measurement simply reveals the corresponding pre-existing state of the system. For the entangled state in Eq. (1), for the possible joint zz spin measurement, the corresponding pre-existing system state is either $|+z\rangle_1|+z\rangle_2$ or $|-z\rangle_1|-z\rangle_2$. If the joint zz spin measurement is performed, the corresponding pre-existing system state is revealed.

The realistic interpretation of entangled state have been successfully utilized to construct a realistic theory that is both local and consistent with the quantum mechanical predictions for the Einstein-Podolsky-Rosen (EPRB) experiment [4] and the Greenberger-Horne-Zeilinger (GHZ) experiment [5]. However, the realistic interpretation of entangled state is still not widely accepted because critics have claimed that the interpretation is flawed in two ways. First, it is claimed that the realistic interpretation cannot distinguish two entangled states which differ only by a relative phase. Second, it is claimed (see [6,7] for example) that, for an ensemble of systems, the realistic interpretation of the entangled state, which mathematically describes each ensemble member, means that the ensemble is interpreted as a mixture of groups of systems where a product state mathematically describes each system in a group. The predictions derived from the mixture are [6-8] generally not the same as those derived from the entangled state and this have led critics to conclude that the realistic interpretation is wrong. In section 2 of this paper, I address these two criticisms and show that they are unfounded.

In spite of the recent wave of proposed local realistic theories [4,5,9-13] which reproduced the quantum mechanical results for the EPRB experiment [4,9-13] and the GHZ experiment [5], there is still a widespread belief [14] that no theory that is simultaneously local and realistic can reproduce the empirically correct quantum predictions for these experiments. Besides these two experiments, there is a third experiment involving a system of two entangled particles, proposed by Lucien Hardy [15], which also tests the predictions of local realistic theory *à la* Bell. Hardy's version of Bell's theorem without inequalities is considered by David Mermin to be the simplest [16] and also the best [17]. Realizations [18-21] of Hardy's experiment with entangled photons also confirmed the quantum predictions instead of the predictions of local realism *à la* Bell. A local realistic theory for Hardy's experiment that can reproduce the quantum mechanical results has not yet been reported. In section 3 of this paper, following my successful approach for the EPRB [4] and GHZ [5] experiments, I present a local realistic theory for Hardy's experiment based on the realistic interpretation of the entangled two-particle Hardy state and show that the theory is consistent with all the probabilistic predictions of quantum theory.

## 2. In defense of the realistic interpretation of entangled state

To address the first criticism, let us consider a system of two spin-½ particles and compare the realistic interpretations of the entangled state in Eq. (1) and the entangled state below which differs only by a relative phase:

$$\frac{1}{\sqrt{2}}|+z\rangle_1|+z\rangle_2 - \frac{1}{\sqrt{2}}|-z\rangle_1|-z\rangle_2. \qquad (2)$$

In both cases, for the possible joint zz spin measurement, the corresponding pre-existing system state is, according to the realistic interpretation, either $|+z\rangle_1|+z\rangle_2$ or $|-z\rangle_1|-z\rangle_2$, i.e., the realistic interpretation does not distinguish the two different entangled states. However, the two entangled states in Eq. (1) and Eq. (2) can be re-expressed in the basis states for a different possible joint spin measurement, for example:

$$\frac{1}{\sqrt{2}}|+z\rangle_1|+z\rangle_2 + \frac{1}{\sqrt{2}}|-z\rangle_1|-z\rangle_2 = \frac{1}{\sqrt{2}}|+x\rangle_1|+x\rangle_2 + \frac{1}{\sqrt{2}}|-x\rangle_1|-x\rangle_2 \qquad (3)$$

and

$$\frac{1}{\sqrt{2}}|+z\rangle_1|+z\rangle_2 - \frac{1}{\sqrt{2}}|-z\rangle_1|-z\rangle_2 = \frac{1}{\sqrt{2}}|+x\rangle_1|-x\rangle_2 + \frac{1}{\sqrt{2}}|-x\rangle_1|+x\rangle_2. \qquad (4)$$

For the possible joint xx spin measurement, the corresponding pre-existing system state is either $|+x\rangle_1|+x\rangle_2$ or $|-x\rangle_1|-x\rangle_2$ in the case of the first entangled state [see Eq. (3)], but the corresponding pre-existing system state is different in the case of the second entangled state [see Eq. (4)]: either $|+x\rangle_1|-x\rangle_2$ or $|-x\rangle_1|+x\rangle_2$. The two different entangled states which differ only by a relative phase are therefore distinguishable in the realistic interpretation, contrary to the first criticism.

To address the second criticism, let us consider an ensemble of two spin-½ particles where each system in the ensemble is mathematically described by the entangled state in Eq. (1). For the possible joint zz spin measurement, the corresponding pre-existing state of each system in the ensemble is, according to the realistic interpretation, either $|+z\rangle_1|+z\rangle_2$ or $|-z\rangle_1|-z\rangle_2$ with equal probability of ½. This means that, from a realistic viewpoint, the corresponding pre-existing state of 50% of the systems in the ensemble is $|+z\rangle_1|+z\rangle_2$ and the corresponding pre-existing state of 50% of the systems in the ensemble is $|-z\rangle_1|-z\rangle_2$. Contrary to the second criticism, the preceding realistic interpretation does not the mean that the ensemble is interpreted as a mixture, where by a mixture it is meant [6-8] that 50% of the systems in the ensemble are each mathematically described by the product state $|+z\rangle_1|+z\rangle_2$ and 50% of the systems in the ensemble are each mathematically described by the product state $|-z\rangle_1|-z\rangle_2$. This is simply because the realistic view "50% of the systems in the ensemble have the pre-existing state $|+z\rangle_1|+z\rangle_2$ ($|-z\rangle_1|-z\rangle_2$)" does not require that these systems have to be mathematically described by the product state $|+z\rangle_1|+z\rangle_2$ ($|-z\rangle_1|-z\rangle_2$).

### 3. A local realistic theory for Hardy's experiment

For simplicity, I will use Mermin's version [16,17] of Hardy's experiment. In this version, a source produces two particles that fly apart in opposite directions, one to the left (*l*) and the other to the right (*r*), each towards a detector. Each detector is set randomly and independently to one of its two only modes (labeled 1 and 2 respectively) before the particles arrive and trigger the detectors. There are only four possible settings for the detectors: 1,1 or 1,2 or 2,1 or 2,2. Each detector flashes either Red (R) or Green (G) when a particle arrives, indicating that the particle is measured R or G respectively.

The source produces a pair of particles that is described by what is called a Hardy state. Following Mermin [16], let us consider the Hardy state given by

$$|\psi\rangle = \sqrt{0.375}|1,R\rangle_l|1,G\rangle_r + \sqrt{0.375}|1,G\rangle_l|1,R\rangle_r - 0.5|1,G\rangle_l|1,G\rangle_r \qquad (5)$$

where [16] for either particle (on the left *l* or right *r*), each of the two eigenstates of observable 1 is defined in terms of the two eigenstates of observable 2 as follows:

$$|1,R\rangle_i = \sqrt{0.4}|2,G\rangle_i + \sqrt{0.6}|2,R\rangle_i, \qquad (6)$$

$$|1,G\rangle_i = \sqrt{0.6}|2,G\rangle_i - \sqrt{0.4}|2,R\rangle_i \qquad i = l \text{ or } r. \qquad (7)$$

The entangled two-particle Hardy state in Eq. (5), which is given in the basis states for the possible 1,1 measurement (i.e., the detector settings are 1,1), can be easily expressed in terms of the basis states for each of the other three possible measurements by using Eq. (6) and Eq. (7):

$$|\psi\rangle = a|1,R\rangle_l|2,G\rangle_r + b|1,G\rangle_l|2,R\rangle_r + c|1,R\rangle_l|2,R\rangle_r, \tag{8}$$

$$|\psi\rangle = d|2,R\rangle_l|1,G\rangle_r + e|2,G\rangle_l|1,R\rangle_r + f|2,R\rangle_l|1,R\rangle_r, \tag{9}$$

$$|\psi\rangle = g|2,R\rangle_l|2,G\rangle_r + h|2,G\rangle_l|2,R\rangle_r + j|2,R\rangle_l|2,R\rangle_r + k|2,G\rangle_l|2,G\rangle_r, \tag{10}$$

where the coefficients are real (their actual values are not important for our discussion) and their squared values, which are the quantum probabilities of joint measurement outcomes, are given in Appendix A in [16]. For example, if the detector settings are 2,2, the probability of measuring G for the left particle and G for the right particle is $k^2$ (see Eq. (10)). If the detector settings are 1,1, the probability of measuring R for the left particle and R for the right particle is 0 (see Eq. (5)).

According to the realistic interpretation of the Hardy state $|\psi\rangle$ in Eqs. (5), (8), (9) and (10), prior to measurement in an experimental run, the system of two particles has four different physical states, one state for each of the four possible detector settings. In particular, for each of the possible detector settings, the corresponding pre-existing system state is one of the basis states $|,\rangle_l|,\rangle_r$ in the expression for $|\psi\rangle$. For example, for the possible 1,1 detector setting, the corresponding pre-existing system state is $|1,R\rangle_l|1,G\rangle_r$ or $|1,G\rangle_l|1,R\rangle_r$ or $|1,G\rangle_l|1,G\rangle_r$. The system of two particles acquires the four system states at the source – these states are determined by 'hidden' variables. Table 1 lists a possible set of four pre-existing system states.

Table 1. A possible set of four pre-existing system states in an experimental run.

| Possible detector setting | Pre-existing system state |
|---|---|
| 1,1 | $|1,R\rangle_l|1,G\rangle_r$ |
| 1,2 | $|1,G\rangle_l|2,R\rangle_r$ |
| 2,1 | $|2,R\rangle_l|1,R\rangle_r$ |
| 2,2 | $|2,R\rangle_l|2,R\rangle_r$ |

From the realistic viewpoint, in an experimental run, measurements merely reveal the corresponding pre-existing system state for the chosen detector settings. For example, if the four pre-existing system states in an experimental run are those in Table 1, the left particle will be measured R and the right particle will be measured G if the detector settings are chosen to be 1,1, revealing the corresponding pre-existing system state $|1,R\rangle_l|1,G\rangle_r$. However, if the detector settings are chosen to be 2,2 instead in the same experimental run, the left particle will be measured R and the right particle will be measured R, revealing the corresponding pre-existing system state $|2,R\rangle_l|2,R\rangle_r$.

In contrast, from the orthodox viewpoint, the system of two particles does not have any definite state before measurement. Measurement on one particle compels that particle to acquire a definite state and instantaneously triggers the other particle, which is spatially separated from the first, to also acquire a definite state. This *instantaneous action-at-a-distance*, i.e., instantaneous triggering of state acquisition over a distance, is, according to Bell [1] and also Mermin [2], *non-locality*. However, from the realistic viewpoint, there is no action-at-a-distance because measurements merely reveal the pre-existing system state corresponding to the chosen detector settings. In other words, my realistic theory for Hardy's experiment is local. My usage of the word 'local' is strictly in keeping with Bell's [1] and Mermin's [2] definition, i.e., local means no action-at-a-distance.

Furthermore, from the realistic viewpoint, for each possible detector setting, the probability that a particular basis state $|,\rangle_l|,\rangle_r$ in the expression for $|\psi\rangle$ is the corresponding pre-existing system state is given by the square of the coefficient for that basis state. Hence the probabilities of joint measurement outcomes are also given by the squared coefficients, in exact agreement with the quantum probabilities. Therefore the local realistic theory presented in this paper for Hardy's experiment reproduces all the quantum probabilistic predictions. Once again, locality and realism can coexist, contrary to conventional belief.